
\magnification=1200
$$\centerline{THE HEAVY FERMION DAMPING RATE PUZZLE}$$\bigskip
\centerline{\bf S. Peigne, E. Pilon and D. Schiff}\smallskip
\centerline{Laboratoire de Physique Theorique et Hautes
Energies\footnote{*}{Laboratoire
associe au Centre National de la Recherche Scientifique}}\par
\centerline{Universite de Paris XI, batiment 211, 91405 Orsay Cedex, France}
\vskip 2cm \baselineskip=20pt
\noindent \underbar{Abstract} \smallskip
We examine again the problem of the damping rate of a moving heavy fermion in a
hot plasma
within the resummed perturbative theory of Pisarski and Braaten. The ansatz for
its evaluation
which relates it to the imaginary part of the fermion propagator pole in the
framework of a
self-consistent approach is critically analyzed. As already pointed out by
various authors,
the only way to define the rate is through additional implementation of
magnetic screening.
We show in detail how the ansatz works in this case and where we disagree with
other authors. We
conclude that the self-consistent approach is not satisfactory.
\vskip 3cm
\noindent LPTHE Orsay 93-13 \par
\noindent April 1993

\vfill \supereject
\noindent {\bf 1. \underbar{Introduction}} \vskip 4mm
Hot gauge theories are the framework to study perturbatively plasmas of weakly
interacting particles. Of special interest are the properties of quasiparticle
modes. The
knowledge of the fermion self-energy, first computed by Klimov and Weldon [1],
has yielded
the dispersion relations of thermal fermionic excitations at leading order in
the coupling
constant g (in hot QED/QCD). They are obtained by equating the real part of the
inverse
effective propagator to zero.\par
To go further in the quasiparticle interpretation, a question has arisen : what
is the
damping rate of this quasiparticle ? In other words, by which coefficient
$\Gamma$ the large
time $e^{\-- \Gamma t}$ behaviour of the corresponding response function is
governed ? When the
effective fermion propagator has a pole, an hypothesis always assumed in
previous works [2-7], the
width $\Gamma$ of the quasiparticle is given by the opposite of the pole
imaginary part.\par
Initially, the damping rate of a gluon at rest raised a problem : one-loop
calculations led to
gauge-dependent results. The solution of this "plasmon puzzle" required to take
into account all
diagrams contributing to leading order in g [3, 8], thus giving a finite and
gauge-independent
result [8, 9].\par Similarly, a finite damping rate $\Gamma$ was found for a
massive fermion at
rest [2]. In that case, only the longitudinal component of the gauge boson
propagator contributes
to $\Gamma$ and the long range interactions are screened by the Debye length in
the framework of
the resummed theory [10]. As soon as the fermion is moving, the transverse
component also
contributes. Its behaviour in the space-like region leads to a logarithmic
infrared divergence [2].
The resummation program is apparently not sufficient to screen the long range
magnetic
interactions. \par Many authors [3-5, 7] have nevertheless tried to calculate
$\Gamma$ with the
hypothesis that the fermion propagator has a pole, i.e. to find the pole
imaginary part. Their
attempts to screen the infrared divergence are based on a self-consistent
approach, originally
proposed by Lebedev and Smilga [3], but results are often different or even
contradictory. Our
purpose here is to study this situation. In agreement with previous authors
[4], we find that
pursuing literally the search of a pole of the propagator (i.e. na\"{\i}vely
using a "complex
energy-shell" condition to define the rate) in the framework of the
self-consistent approach leads
to a failure : the logarithmic singularity is not screened. On the other hand,
going on with this
approach in the QCD case, in presence of magnetic screening, leads to an answer
for the rate only
in the case of a fast moving fermion ; the corresponding equation for a non
relativistic heavy
fermion has no solution. The result obtained for the relativistic fermion
differs from the usual
intuitive "narrow width" expression, $O(g^2 T \ell n(1/g))$ pointing out at
possible difficulties
connected with the simple minded analytic continuation performed to define the
propagator in the
complex energy plane. In this respect, let us stress that our results disagree
with those of ref.
[7]. We think that the source of the disagreement lies in the way analytic
continuation is
performed to evaluate the damping rate.\par  After relating the damping rate
with the presumed pole
of the fermion propagator (section 2), we calculate the fermion self-energy in
the imaginary time
formalism and perform the analytic continuation to complex external energy
(section 3). In section
4, first discussing the solution obtained without magnetic screening, we then
analyze the QCD case
in presence of a magnetic mass both for a nonrelativistic and a fast heavy
fermion. Section 5 is the
conclusion. \vskip 5mm \noindent {\bf 2. \underbar{Definition of the damping
rate}} \vskip 4mm We
assume the existence of a pole in the complete fermion propagator. The damping
rate is then given
by the opposite of the imaginary part of this pole. Working in the imaginary
time formalism, the
complete fermion propagator is :

$$S(p_4, \vec p) = {1 \over - i \not p + M - \Sigma_E(p_4, \vec p)} \ \
,\eqno(1)$$

\vfill \supereject
\noindent where $\Sigma_E$ is the fermion self-energy in euclidean
space\footnote{*}{In the
imaginary time formalism, the euclidean Dirac algebra $\lbrace \gamma^{\mu} ,
\gamma^{\nu} \rbrace
= 2 \delta^{\mu \nu}$ is used. We note ${\not p} = p_4 \gamma_0 + \vec p \ \vec
\gamma$, where
$p_4$ is the discrete euclidean coordinate $p_4=(2n+1) \pi T$. The continuation
for
$\gamma$-matrics is achieved by $i \vec \gamma \to - \vec \gamma$ ($\gamma_0$
unchanged) to recover
the algebra $\lbrace \gamma^{\mu},\gamma^{\nu} \rbrace = 2g^{\mu \nu}$.}. The
usual analytic
continuation for the retarded propagator is $ip_4 \to p_0 + i \varepsilon$ with
$p_0$ real. In
order to locate the position of the complex pole in the lower half energy
plane, we tentatively
assume that we may continue the retarded propagator to complex values of $p_0$
as $ip_4 \to p_0$.
The pole is the value of $p_0$ for which :

$$det \left\{ p_0 \gamma_0 - \vec p \ \vec \gamma - M - \Sigma(p_0, \vec p)
\right\} = 0 \ \ , \eqno(2)$$

\noindent where $\Sigma(p_0, \vec p) = - \Sigma_E(- i p_0, \vec p)$. In
general, $p_0$ is
complex :

$$p_0 = E - i \Gamma \ \ ,\eqno(3)$$

\noindent with $\Gamma$ the fermion damping rate. \par
By rotational invariance in the rest frame of the plasma, $\Sigma$ takes the
form :

$$\Sigma = a \gamma_0 + b \vec p \ \vec \gamma + c \ \ ,\eqno(4)$$

\noindent which from eq. (2) leads to :

$$p_0 = a \pm \sqrt{(1 + b)^2 \vec p^{\ 2} + (M + c)^2} \ \ . \eqno(5)$$

In eq. (5), a, b and c are functions of $p_0$ so that eq. (5) is a
transcendental equation,
yielding the pole $p_0 = E - i \Gamma$ as an implicit function of $|\vec p|$.
\par
For a heavy fermion $(M >> T)$, $a << \sqrt{\vec p^{\ 2} + M^2}$ and ${\rm Re}
\ p_0 \equiv E
\simeq \sqrt{\vec p^{\ 2} + M^2}$. We use the perturbative approximations $b <<
1$, $c << M$, and
neglect quadratic terms in $b$ and $c$. Then we show that :

$$\left.\Gamma = - {\rm Im} \ p_0 \simeq - {1 \over 4E} {\rm Im} \ Tr (\not p +
M) \Sigma(p_0, \vec
p) \right|_{p_0= E - i \Gamma} \eqno(6)$$

\noindent where ${\not p} = E \gamma_0 - \vec p \ \vec \gamma$.
Notice that $\Gamma$ is determined by a functional equation since $\Sigma$ is
evaluated at $p_0
= E - i \Gamma$. In principle, going off the real energy axis requires the
knowledge of the
analytic structure of $\Sigma$, namely the existence and positions of cuts if
any, as well as
the different sheets, especially the one on which the pole is to be searched.
\par
Since this analytic structure is largely unknown, it may be dangerous to
explore the complex
plane naively without any guidance, as will be discussed in sect. 3. \par
\vskip 5mm
\noindent {\bf 3. \underbar{Calculation of Im $\Sigma$ for a complex external
energy}}\vskip 4mm
The leading contribution to the hard fermion self-energy in the imaginary time
formalism reads :

$$\Sigma (p) = - g^2 C_F \int_{soft} {d^3q \over (2 \pi)^3} \ T \sum_{q_4=2 \pi
n T} \left\{
\gamma_{\mu} S_0(p') \gamma_{\nu} \ ^{*}\Delta^{\mu \nu}(q) \right\} \eqno(7)$$

\noindent $C_F$ is the QCD Casimir which should be replaced by 1 in the case of
QED. $S_0(p')$ is
the bare internal fermion propagator ($p' = p - q$) :

$$S_0(p') = {i \not {p'} + M \over p_4^{'2} + E'^2} \quad {\rm with} \quad E'^2
=
\overrightarrow{p '}^2 + M^2 \eqno(8)$$

\noindent *$\Delta^{\mu \nu}(q)$ is the resummed gauge boson propagator
including hard thermal
loops [11]. In the Coulomb gauge, it consists of two terms : \par
$\bullet$ the longitudinal part *$\Delta^{00}$ = *$\Delta_{\ell}(q)$. \par
$\bullet$ the transverse part *$\Delta^{ij}$ = $\widehat{Q}^{ij}$
*$\Delta_t(q)$, where
$\widehat{Q}^{ij} = \delta^{ij} - \widehat{q}^{ \ i} \widehat{q}^{ \ j}$, with
$\widehat{q}_i =
{q_i \over |q|}$. \par \noindent *$\Delta_{\ell}$ gives a finite contribution
to ${\rm Im} \
\Sigma$, whereas *$\Delta_t$ is responsible for infrared problems [2].
Therefore we shall focus on
the contribution which comes from the transverse part :

$$\Sigma_t = - g^2 C_F \int_{soft} {d^3 q \over (2 \pi)^3} T \sum_{q_4} \
^{*}\Delta_t(q_4,\vec q)
\left\{ \gamma^i S_0(p') \gamma^j \widehat{Q}^{ij} \right\} \ \ . \eqno(9)$$

\noindent In order to sum over $q_4$, we use the spectral density
representation for the
transverse propagator :

$$^{*}\Delta_t(q_4, \vec q) = \int_0^{\beta} d\tau e^{iq_4 \tau} \int_{-
\infty}^{\infty} d
\omega \ \rho_t(\omega, q) \ (1 + n(\omega)) \ e^{-\omega \tau} \ \
,\eqno(10)$$

\noindent where $n$ denotes the Bose-Einstein statistical factor and the
transverse spectral
density $\rho_t(\omega, q)$ given in [11] is proportional to the imaginary part
of
*$\Delta_t(q_4, \vec q)$ after the continuation $iq_4 \to \omega + i
\varepsilon$ :

$$\rho_t(\omega, q) = {1 \over \pi} {\rm Im} ^{*}\Delta_t (q_4 \to - i(\omega +
i \varepsilon),
\vec q) \ \ . \eqno(11)$$

\noindent To calculate ${\rm Im} \ \Sigma$ for a complex external energy we may
use a mixed
representation for the bare fermion propagator :

$$S_0(p) = \int_0^{\beta} d \tau \  e^{ip_4 \tau} \lbrack \tilde n(- E) \ e^{-E
\tau} P_+ +
\tilde n(E) \ e^{E \tau} P_- \rbrack \eqno(12)$$

\noindent $\tilde n$ refers to the Fermi-Dirac statistical factor and
$P_+(P_-)$ project on
positive (negative) energy states :

$$P_{\pm} = {1 \over 2E} \lbrack E \gamma_0 \pm (i \vec p \ \vec \gamma + M)
\rbrack \ \ .
\eqno(13)$$

\noindent Instead of using eq. (12), Pisarski [7] proposes an ansatz for the
fermion
propagator in the spirit of eq. (10) : the fermion spectral density is taken as
a Breit-Wigner
distribution. This however does not allow exploring the pole position in the
lower half
complex plane. The analytic continuation which is used in [7] in order to go
from the evaluation
of ${\rm Im} \ \Sigma$ on the real axis to its value on the complex pole seems
quite questionable to
us as we shall argue in section 4. \par We insert eqs. (10) and (12) in eq. (9)
and find after the
proper analytic continuation (keeping only the soft space-like region
$|q_0| < q << T$) :

$${\rm Im} \ \Sigma(p_0, \vec p) = 2g^2 C_F T \int_{soft} {d^3q \over (2
\pi)^3} \int_{-q}^{+q}
{dq_0 \over q_0} \rho_t(q_0, q) {\rm Im} \ {p'_0 \gamma_0 - (\vec p \ '.
\widehat {\vec q})
(\widehat {\vec q}. \vec \gamma ) - M \over p'^{2}_{0} - E'^2} \eqno(14)$$

\noindent which is the result found in [4] with the real time formalism. We
have used $|q_0| << T <<
E'$ and $n(q_0) \simeq {T \over q_0}$. Moreover, as $p_0$ is hard, $|p_0| \sim
E$, we replace $p'_0
= p_0 - q_0$ by $E$ in the numerator of eq. (14). We also write :

$${1 \over p'^{2}_{0} - E'^2} = {1 \over 2p'_0} \left( {1 \over p'_0 - E'} + {1
\over p'_0 +
E'} \right) \simeq {1 \over 2E} \ {1 \over p'_0 - E'} \ \ , \eqno(15)$$

\noindent as $E$ and $E'$ are hard energies. We get :

$${\rm Im} \ \Sigma(p_0, \vec p) = g^2 C_F {T \over E} \int_{soft} {d^3q \over
(2 \pi)^3}
\int_{-q}^{+q} {dq_0 \over q_0} \rho_t (q_0, q) \left( E \gamma_0 - (\vec p \ '
\ \widehat{\vec q})
(\widehat{\vec q} \ \vec \gamma) - M \right) {\rm Im} \ {1 \over p'_0 - E'}
\eqno(16)$$

{}From eqs. (6) and (16) we obtain :

$$\Gamma(p) = {g^2 C_FT \over 4 \pi^2} v^2 \int_0^{q*} dq \ q^2 \int_{-q}^{+q}
{dq_0 \over q_0}
\rho_t(q_0, q) \int_{-1}^1 dx (1 + {q \over p} x - x^2) {\rm Im} \ {- 1 \over
p'_0 - E'} \eqno(17)$$

\noindent where $v = {p \over {\rm E} }$ is the fermion velocity and $x = \cos
\theta$ with $\theta$
the angle between $\vec q$ and $\vec p$. The cut-off $q^{*}$ determining the
soft integration
region is some arbitrary scale $<< T$ e.g. $T \sqrt{g}$. \par
\vskip 5mm
\noindent {\bf 4. \underbar{Calculation of the damping rate}} \par
\vskip 4mm
In order to compute $\Gamma(p)$, the "narrow width" approximation was used as a
first attempt ;
as in the case of a fermion at rest, the external energy was taken to be real.
But this led to an
infrared logarithmic divergence [2]. In order to cure this pathology, Lebedev
and Smilga
[3] proposed a self-consistent approach according to which the non-zero value
of $\Gamma$ itself
would screen this divergence. The practical implementations of this philosophy
differ however in the
literature. Here we first follow the procedure of Baier, Nakkagawa and Niegawa
[4] which allows
us to deal with real or complex external energies. We now restrict ourselves to
$p \geq T$ and
neglect the ${q \over p} x$ term in eq. (17). \par
The self-consistent approach amounts to introduce the rate $\Gamma$ as a
dissipative coefficient in
the bare fermion propagator [12], so that in eq. (8), $E'$ is replaced by
$\widehat{E}'$ :

$$\widehat{E}' = E' - i \Gamma(p') \ \ . \eqno(18)$$

\noindent In other words, the only singularity of the effective dissipative
propagator is the
complex pole at $E - i \Gamma$. With these modifications, eq. (17) becomes :

$$\Gamma(p) = {g^2C_FT \over 4 \pi^2} v^2 \int_0^{q*} dq \ q^2 \int_{-q}^q
{dq_0 \over q_0}
\ \rho_t(q_0, q) \ J(q_0, q)$$
\noindent with
$$J(q_0, q) = \int_{-1}^1 dx (1 - x^2) {\rm Im} \left({- 1 \over p'_0 -
\widehat{E}'} \right) \ \ .
\eqno(19)$$

\noindent Approximating $\Gamma(p') \simeq \Gamma(p)$, one finds [4] :

$$\Gamma(p) = {g^2 C_F T \over 2 \pi^2} \ v \int_0^{q*} dq \ q \int_{-q}^q
{dq_0 \over q_0}
\rho_t(q_0, q)$$
$$\left\{arctan \left({{\rm Re} \ p_0 - E + vq - q_0 \over {\rm Im} \ p_0 +
\Gamma(p)} \right) -
arctan \left( {{\rm Re} \ p_0 - E - vq - q_0 \over {\rm Im} \ p_0 +
\Gamma(p)}\right) \right\} \
\ .  \eqno(20)$$

\vskip 3mm
\noindent To apprehend the infrared behaviour of the r.h.s. of eq. (20), it is
sufficient to put
$q_0 = 0$ in the curly bracket, and to use the limiting form of $\rho_t(q_0,
q)$ in the space-like
region [11]
$$\rho_t (q_0, q) \ {= \atop{q_0 << q << m_g} } \ {1 \over q^2} \  {1 \over
\pi}
{\left( {3 \pi \over 4} \ m_g^2 {q_0 \over q^3} \right) \over 1 + \left( {3 \pi
\over 4}
m_g^2 {q_0 \over q^3} \right)^2} \ \ ,  \eqno(21)$$

\noindent where $m_g$ is the thermal mass of the gauge boson. We get :
\vfill \supereject
$$\Gamma(p) = {g^2 C_F T \over 2 \pi^2} v \int_0^{q*} {dq \over q} \ {2 \over
\pi} \ arctan
\left( {3 \over 4} \pi {m_g^2 \over q^2} \right) .$$
$$. \left\{arctan {{\rm Re} \ p_0 - E + vq \over {\rm Im} \ p_0 +
\Gamma(p)} - arctan {{\rm Re} \ p_0 - E - vq \over {\rm Im} \ p_0 + \Gamma(p)}
\right\} \ \ .
\eqno(22)$$

\vskip 5mm
${\bf a) \underline{Absence \ of \ magnetic \ screening \ (QED \ case)} }$ \par
\vskip 4mm
As noticed in [4], with $p_0 = E - i \Gamma$, which is required in principle by
eq. (6), the
r.h.s. of eq. (22) presents a singularity. Thus the self-consistent approach
alone does not
achieve the initial aim which was to screen the logarithmic divergence. In fact
the origin of
this divergence is well-known : it is due to the ${1 \over q^2}$ factor in eq.
(21) which
reflects the absence of magnetic screening in the static limit at this order.
It may seem
surprising to try to cure a bosonic disease by acting on the fermionic sector,
as in the
self-consistent approach, and the success of such a treatment in the screening
of the
divergence would have been somewhat miraculous. \par
Let us stress that the results found in [5] and [7] for the non-relativistic
fermion by
putting ${\rm Im} \ p_0 = 0$ in the r.h.s. of eq. (22) do not in principle give
the damping rate,
since this procedure deviates from the prescription of eq. (6). Although a
dissipative coefficient
is introduced, the choice ${\rm Im} \ p^0 = 0$ is a kind of reminiscense of the
narrow width
approximation, the validity of which is precisely questioned by the divergence
at the pole :
thus the link with the damping rate is unclear and the finiteness of the
resulting expressions
may be misleading. \par
Strictly speaking, the r.h.s. of eqs. (20), (22) rely on the replacement
$\Gamma(p') \to
\Gamma(p)$ in eq. (19). Since $\Gamma$ is not independent of $p$, this
approximation is correct as
long as $|\Gamma(p') - \Gamma(p)| << |{\rm Im} \ p^0 + \Gamma(p)|$, but it
fails at ${\rm Im} \ p^0
= - \Gamma(p)$, so that the conclusion of [4] is not rigorously stated. Since
$\Gamma(p') -
\Gamma(p) \sim O(q)$ the crude replacement of ${\rm Im} \ p_0 + \Gamma(p)$ by
this $O(q)$ term in
eq. (22) still does not cure the divergence at the pole : the conclusions of
[4] are expected to
remain unaltered. \par \vfill \supereject We will show however that the
situation worsens : we
obtain a functional equation for $\Gamma(p)$ which has no solution even in
presence of magnetic
screening (at least in the non relativistic case) ! Using

$$\Gamma(p') \simeq \Gamma(p) - q \ {\partial \Gamma \over \partial p}\ x \ \ ,
\eqno(23)$$

\noindent which holds\footnote{*}{Eq. (23) assumes the differentiability of
$\Gamma(p)$ with respect to $p$. This
assumption may seem quite restrictive and not founded. Let us notice however
that : \par
\item{1 -} it is natural in the framework of the pole ansatz which supposes
that the pole is
isolated from other singularities.\par \item{2 -} it is less stringent than
assuming $\Gamma(p')
= \Gamma(p)$.\par
\item{3 -} Finally it is more consistent than the hypothesis of constant
behaviour, which eventually
leads [5, 7] to $\Gamma(p)$ being a differentiable function of $v(p) =
p/E(p)$.\par} if $\left| {q
\over \Gamma} \ {\partial \Gamma \over \partial p} \right| << 1$, and with $EE'
= E + {q^2 \over
2E} - v q x$, the function $J$ in Eq. (19) is given by :

$$J(q_0, q) = \int_{-1}^1 dx {(1 - x^2) ({\rm Im} \ p_0 + \Gamma(p) - q
{\partial \Gamma \over
\partial p} x) \over \lbrack {\rm Re} \ p_0 - q_0 - E - {q^2 \over 2E} + vq x
\rbrack^2 + \lbrack
{\rm Im} \ p_0 + \Gamma(p) - q {\partial \Gamma \over \partial p} \ x \rbrack^2
} \ \ .  \eqno(24)$$

\vskip 3mm
\noindent Putting $p_0 = E - i \Gamma$, and after some simple manipulations we
find (in the
perturbative approximation ${\partial \Gamma \over \partial p} << v)$ that J
may be safely
approximated by :

$$J = - sgn \left( {\partial \Gamma \over \partial p} \right) \ sgn \left( {q_0
\over q} + {q \over
2E} \right) {1 \over qv} \ \pi \ \theta \left( v - \left| {q_0 \over q} + {q
\over 2E} \right|
\right)
 \ \ . \eqno(25)$$

\noindent Then eq. (19) gives :

$$\Gamma (p) = - sgn \left( {\partial \Gamma \over \partial p} \right) {g^2 C_F
T \over 4 \pi} v
\int_0^{q*} dq \ q \int_{-q}^q {dq_0 \over q_0} \rho_t(q_0, q) sgn \left( {q_0
\over q} + {q \over
2E} \right) \theta \left( v_- \left| {q_0 \over q} + {q \over 2{\rm E} }
\right| \right)
\eqno(26)$$

\noindent As far as the QED case is concerned (absence of magnetic mass) the
behaviour of
$\rho_t(q_0, q)$ in the space-like static limit (cf. eq. (21)) still generates
a logarithmic
singularity.

\vskip 5mm
${\bf b) \underline{Presence \ of \ magnetic \ screening \ (QCD \ case)} }$
\par
\vskip 4mm
The unexpected novelty is that the occurrence of magnetic screening does not
seem to solve the
fermion damping puzzle, at least in the non-relativistic regime ! \par
Unfortunately we do not know the spectral function $\rho_t(q_0, q)$ in presence
of a magnetic
mass, except in the static limit. In order to explore the effect of magnetic
screening, we
propose a simple parametrization which interpolates smoothly between : \par
$\bullet$ the static limit, where we expect the magnetic "mass" $\mu$ to screen
the infrared
singularity : ${\rm Re} \ \Pi_t(q_0, q)$ $\longrightarrow \atop{{|q_0|<<q}\atop
{q_0,q\to 0} }$
$\mu^2$ so that $\rho_t(q_0, q)$ $\sim \atop{{q_0\to 0} \atop{q<<m_g}}$ ${1
\over q^2 + \mu^2}$,
instead of ${1 \over q^2}$ as in eq.222222221). \par
$\bullet$ the region away from the static limit, where the hard-loop
approximation of
$\rho_t(q_0, q)$ certainly holds. Hence we take :

$$\rho_t^{model}(q_0, q) = {{3 \over 4} m_g^2 {q_0 \over q} \over (q^2 +
\mu^2)^2 + \left( {3 \over
4} \pi m_g^2 \right)^2 \left( {q_0 \over q} \right)^2} \ \ . \eqno(27)$$

\noindent This leads to :

$$\Gamma(p) = - sgn \left( {\partial \Gamma \over \partial p} \right) {g^2 C_F
T \over 4 \pi^2} v
\int_0^{q*} dq \ {q \over q^2 + \mu^2} \ \ .$$
$$ . \left\{ 2 \ arctan \left( { {q \over 2E} \nu \over q^{2} + \mu^{2}}
\right) + arctan \left(
{ \left( v - {q \over 2E} \right) \over q^2 + \mu^2} \nu \right) - arctan
\left( { \left( v + {q
\over 2E} \right) \over q^2 + \mu^2} \nu \right) \right\} \eqno(28)$$

\vskip 3mm
\noindent where $\nu = {3 \over 4} \pi m_g^2$. \par
One can show that the curly bracket in eq. (28) is dominated by the first term
yielding:

$$ \Gamma(p) = - sgn \left( {\partial \Gamma \over \partial p} \right) {g^2 C_F
T \over 2 \pi^2} v
\int_0^{\infty} dr {r \over r^2 + 1} arctan \left( K {r \over r^2 + 1} \right)
\ \ , \eqno(29)$$

\vskip 3mm
\noindent where we have set $r = {q \over \mu}$ and $K = {3 \over 8} \pi {m_g^2
\over
{\rm E} \mu}$. The $q^{*}$ dependence of eq. (28) is subleading in $g$ and can
be dropped. \par
The result is :

$$\Gamma(p) = - sgn \left( {\partial \Gamma \over \partial p} \right) {g^2 C_F
T \over 4 \pi} v
\ sinh^{-1} \left[ {3 \pi \over 16} {m_g^2 \over {\rm E} \mu} \right] \ \ .
\eqno(30)$$

\vskip 3mm
Consequently $\Gamma(p)$ should satisfy the condition :

$$\left| {\partial \Gamma \over \partial p} \right| = - {g^2 C_F T \over 4 \pi}
{\partial \over
\partial p} \left\{ {p \over \sqrt{p^2 + M^2}} sinh^{-1} \left( {3 \pi \over
16} {m_g^2 \over \mu
\sqrt{p^2 + M^2} } \right) \right\} \ \ . \eqno(31)$$

\vskip 3mm
\noindent When the expression between brackets is an increasing function of
$p$, eq. (31) has
no solution. This happens when $p < M$. Thus there is no solution for a
non-relativistic
fermion: the corresponding propagator has no pole. In the case where $v << 1$,
the $q_0$
integration range is restricted to the static limit region $q_0 << q$ where the
parametrization (27) is certainly reliable. When $ p \geq M$, eq. (31) has a
solution.
Namely, in the ultrarelativistic case we get :

$$\Gamma(p)\left|_{v=1} \right. = {g^2 C_F T \over 4 \pi} sinh^{-1} \left[ {3
\pi \over 16} {m_g^2
\over {\rm E} \mu} \right] \ \ . \eqno(32)$$

\noindent Assuming $\mu \sim g^2T$, ${m_g^2 \over {\rm E} \mu} \sim {T \over E}
<< 1$, we thus
have :

$$\Gamma(p)\left|_{v=1} \right. = {3g^2 C_F T \over 64} \ {m_g^2 \over {\rm E}
\mu} \eqno(33)$$ \
\

Surprisingly, we do not find the logarithmic dependence in $\mu$ naively
expected since $\mu$
screens the logarithmic divergence. We may first question the reliability of
the
parametrization (27) used in the whole $q_0$ integration range. However, due to
the sign flip
of $J(q_0, q)$ and the odd property of $\rho_t(q_0, q)$ under the change $q_0
\to - q_0$, the
essential part of the $q_0$ integral in eq. (26) comes from the narrow window
$\lbrack - {q
\over 2E} ; {q \over 2E} \rbrack$ where the ansatz (27) is expected to hold
\footnote{*}{The
second window let open by $J$ and the odd behaviour of $\rho_t$, which is $v -
{q \over 2E} \leq
{q_0 \over q} \leq v + {q \over 2E}$, gives only a negligible contribution with
respect to
the one we discuss.}. Then the nature of the $\mu$-dependence of $\Gamma(p)$
(for $p \geq
M$) is controlled by the ratio of the width of the window $({q \over E})$ to
the width of
the spectral density $\rho_t^{ model}$ considered as a function of ${q_0 \over
q}$ which is $({q^2 +
\mu^2 \over {3 \over 4} \pi m_g^2})$. The magnitude of this ratio $R$ given by
:

$$R = K {2 r \over r^2 + 1} \ \ , \eqno(34)$$

\noindent with $r = {q \over \mu}$, is controlled by the parameter $K$ defined
in eq. (29).
\par
If $\mu$ is arbitrarily small, namely $\mu << {m_g^2 \over E} << O(g^2T)$ the
spectral
density $\rho_t$ is sharply peaked. It is then legitimate to replace as done in
previous works
[4], $q_0$ by $0$ in the integrand of eq. (26), except in ${1 \over q_0} \
\rho_t(q_0, q)$. In
this case, instead of (33), eq. (32) yields indeed the logarithmic dependence
on $\mu$ :

$$\Gamma(p)\left|_{v=1\atop\mu \to 0} \right. = {g^2 C_F T \over 4 \pi} \ell n
\left( {3 \pi
\over 8} {m_g^2 \over {\rm E} \mu} \right) \eqno(35)$$

\noindent But on the other hand, for the physically sensible value $\mu \sim
O(g^2 T)$, $K <<
1$ and the density $\rho_t$ has a broader support than the window $- {q \over
2E} \leq {q_0 \over
q} \leq {q \over 2E}$. This forbids the appearance of the logarithmic
dependence and leads to
eq. (33). \par
We point out that the sign flip in $J$, hence the narrowness of the window ${q
\over E}$
instead of $O(1)$, are a direct consequence of the self-consistent approach
associated with
the pole prescription of eq. (6), through eqs. (23)-(24). If one would instead
use a nondissipative
fermion and work with a real energy $p^0$ - as usual in the narrow width
approximation -, $J$
would be replaced by the integral over $\cos \theta$ of a $\delta$-function,
which would not produce
such a sign flip. In such a case the corresponding ratio $R$ would be
controlled by $m_g^2/\mu^2$,
much larger than 1. One would then get the standard $\ell n(m_g/\mu)$
dependence in $\Gamma$.
This remark leads to the conclusion that even in presence of magnetic screening
and when the
pole exists, the narrow width approximation and the self-consistent approach
disagree. \par
This unconventional conclusion, associated with the qualitative difference
between the
nonrelativistic and ultrarelativistic regimes without any apparent physical
reason make us feel
uneasy. Hence we wonder about the correctness of the self-consistent procedure,
i.e. the pole
ansatz of the dissipative propagator together with the exploration of the lower
half complex
energy plane, irrespectively of the true analytic structure of the complete
fermion propagator.
\par
In ref. [7], arguments are given to obtain informations about the singularity
structure of the
propagator and discuss the possibility of ignoring this structure, keeping only
the pole. We
do not, however, think that these arguments are well founded. In fact, the
analytic
continuation which is used in ref. [7], in order to explore the lower
half-plane of the energy,
is performed on the self-energy imaginary part evaluated for real external
energies. But,
clearly :

$${\rm Im} \ \Sigma({\rm real} \ p_0 \to {\rm complex} \ p_0) \not= {\rm Im} \
\left( \Sigma ({\rm
real} \ p_0 \to {\rm complex} \ p_0) \right)$$

\noindent and it is only the latter quantity which is of interest and which we
actually study in the
present work.\par
In fact, instead of eq. (12), ref. [7] uses a spectral density representation
for
the internal fermion propagator, which is correct only for a real external
energy $p_0$. To take
into account the width of the fermion, a Breit-Wigner form for the fermion
spectral density is introduced by hand. To make contact with this method, we
write the imaginary
part of eq. (15) with a dissipative energy $\widehat{E}'$ as :

$${\rm Im} \ {1 \over p'_0 - \widehat{E}'} = - {\Gamma(p') + {\rm Im} \ p_0
\over (E - E' - q_0)^2 +
(\Gamma(p') + {\rm Im} \ p_0)^2} \eqno(36)$$

\vskip 3mm
\noindent We recover the same Breit-Wigner form as ref. [7] only for ${\rm Im}
\ p_0 = 0$.
The ad hoc analytic continuation which is proposed in order to go to ${\rm Im}
\ p_0 = - \Gamma$
leads to results contradicted by our treatment, which enables us to go directly
to the pole. In
particular, this is flagrant for the case of a non-relativistic fermion for
which ref. [7] states
that no magnetic screening is needed. Also when going to the complex pole in
eq. (36), one finds as
a consequence that the final answer for $\Gamma(p)$ cannot depend on itself as
appears in [7] but
only on $\Gamma(p') - \Gamma(p)$ which is what we have dealt with. \par

\vskip 5mm
\noindent {\bf 5. \underbar{Conclusion}} \vskip 4mm
By a careful exploration of the location of the pole of the fermion propagator,
we have shown
that the self-consistent approach using a dissipative fermion is of marginal
relevance to solve
the heavy moving fermion damping puzzle : \par
$\bullet$ it does not provide any screening for the infrared divergence, which
has to be cured
by magnetic screening. \par
$\bullet$ In presence of magnetic screening, the situation depends on the
kinematical regime.
\par
In the non relativistic regime the functional equation obtained for $\Gamma$
has no  solution.
The situation changes dramatically in the relativistic regime, but without any
apparent
physical reason : we do find a pole in this case, hence a value for $\Gamma$.
However we got a
rather unusual result, especially concerning the dependence of $\Gamma$ on the
magnetic mass
$\mu$, as compared to the commonly expected $\ell n \ \mu$. The
$\mu$-dependence which we get in
$\Gamma$ is logarithmic only if $\mu$ is very small with respect to the
standard scale $O(g^2
T)$ and this is a direct and specific consequence of the use of a dissipative
fermion. This
disagreement with the result obtained in the narrow width approximation (which
is naively
expected to work since $\Gamma \sim g^2 T << gT << M$) suggests that the
self-consistent
approach is not a safe procedure and may lead to question the validity of the
simple hypothesis of
a leading pole singularity. The complete singularity of the propagator is
unknown : there are no
physical requirements to determine it and in particular no way to define a
physical sheet as for
the $T = 0$ $S$-matrix element. \par
\vfill \supereject
Notice that the absence of a pole in the fermion propagator
does not jeopardize, in principle, the quasi-particle interpretation of plasma
excitations. Indeed,
a quasi-particle shows up in the large time asymptotic behaviour of the mixed
time/momentum
representation of the (retarded) fermion Green's function as :

$$G(t, p)  _{{\sim} \atop{t\to + \infty} } A(t, p) \ exp \lbrack - i
\omega_0(p)t - \Gamma(p)t
\rbrack$$

\noindent where $\omega_0(p)$ gives the dispersion relation and $\Gamma(p)$ is
the damping
rate. In the complex energy plane, $p_0 = \omega_0(p) - i \Gamma(p)$ is the
nearest
singularity of the propagator with respect to the real axis, as can be shown,
e.g. by a
saddle point approximation, and this singularity need not be a pole, but e.g. a
branch point as
claimed by Smilga [13] in the QED case. The prefactor A is non exponential in
time and depends on
the nature of this singularity (for example it is a constant for a pole). In
practice however,
locating any singularity off the real energy axis (a pole, a branch point or an
essential
singularity) involves the knowledge of the analytic structure of the finite
temperature propagator
which is out of reach at present. \par

\vskip 5mm
\noindent ${\bf \underline{Acknowledgments}}$ \par
We wish to thank R. Baier and A. Smilga for useful discussions, as well as T.
Altherr and R.
Pisarski for stimulating correspondence. \par

\vfill \supereject
\centerline{\bf \underbar{References}} \par
\vskip 5mm
\item{[1]} V. V. Klimov, Yad. Fiz. $\underline{33}$ (1981) 1734 ; [Sov. J.
Nucl. Phys.
$\underline{33}$ (1981) 934]. \par
\item{} H. A. Weldon, Phys. Rev. $\underline{D26}$ (1982) 2789. \par
\item{[2]} R. Pisarski, Fermilab-Pub-88/123-T (09/88) ; Phys. Rev. Lett.
$\underline{63}$
(1989) 1129. \par
\item{[3]} V. V. Lebedev and A. V. Smilga, Ann. Phys. (N. Y.)
$\underline{202}$, 229 (1990)                           Phys.
Lett. $\underline{253B}$, 231 (1991) ; Physica $\underline{A181}$, 187 (1992).
\par
\item{[4]} R. Baier, H. Nakkagawa, A. Niegawa, Osaka University preprint
OCU-PHYS-145 to be
published in the Canadian Journal of Physics. \par
\item{[5]} T. Altherr, E. Petitgirard and T. del
Rio Gaztellurutia, Phys. Rev. $\underline{D47}$ (1993) 703. \par
\item{[6]} R. Kobes, G. Kunstatter
and K. Mak, Phys. Rev. $\underline{D45}$ (1992) 4632. \par
\item{[7]} R. D. Pisarski, preprint
BNL-P-1/92. \par
\item{[8]} E. Braaten, R. D. Pisarski, Phys. Rev. $\underline{D42}$ (1990)
2156.
\par
\item{[9]} R. Baier, G. Kunstatter and D. Schiff, Phys. Rev. $\underline{45}$
(1992) 4381 ; Nucl. Phys. $\underline{B388}$ (1992) 287 ; A. Rebhan, CERN
preprint,
CERN-TH/6434/92. \par
\item{[10]} E. Braaten, R. D. Pisarski, Nucl. Phys. $\underline{B337}$ (1990)
569 ; Phys. Rev. Lett. $\underline{64}$ (1990) 1338. \par
\item{[11]} R. D. Pisarski, Physica $\underline{A158}$ (1989) 146. \par
\item{[12]} See for instance : I. Hardman, H. Umezawa and Y. Yamanaka, J. Math.
Phys.
$\underline{28}$ (1987) 2925. \par
\item{[13]} A. V. Smilga, Bern University preprint BUTP-92/39. \par

\bye